\documentclass[twocolumn,a4paper]{article}
\usepackage{graphics}
\usepackage{epsfig}

\begin{document}
\title{Coupling of two superconductors through a ferromagnet: evidence for a $\pi$-junction}
\author{V. V. Ryazanov$^1$, V. A. Oboznov$^1$, A. Yu. Rusanov$^1$,
A. V. Veretennikov$^1$,\\  A. A. Golubov$^2$, and J. Aarts$^3$ \\
$^1$* Institute of Solid State Physics, Russian Academy of
Sciences, \\ Chernogolovka, 142432, Russia \\ $^2$ University of
Twente, P. O. Box 217, 7500 AE Enschede, the Netherlands \\ $^3$
Kamerlingh Onnes Laboratory, Leiden University, \\ P.O. Box 9504,
2300 RA Leiden, the Netherlands}

\date{\today}

\maketitle

\begin{abstract}
We report measurements of the temperature dependence of the
critical current in Josephson junctions consisting of conventional
superconducting banks of Nb and a weakly ferromagnetic interlayer
of a Cu$_x$Ni$_{1-x}$ alloy, with $x$ around 0.5. With decreasing
temperature $I_c$ generally increases, but for specific
thicknesses of the ferromagnetic interlayer, a maximum is found
followed by a strong decrease down to zero, after which $I_c$
rises again. Such a sharp cusp can only be explained by assuming
that the junction changes from a 0-phase state at high
temperatures to a $\pi$-phase state at low temperatures.
\end{abstract}

\section*{}
Almost all of the presently known superconductors contain
conventional Cooper pairs, two electrons with opposite spin and
momentum (+{\bf k}$\uparrow$,-{\bf k}$\downarrow$). Such a system
is described by an isotropic excitation gap and / or order
parameter. The exceptions are found notably in high $T_c$ (oxide)
superconductors, and in some Heavy Fermion systems, in which cases
the exact pairing mechanism is not yet fully understood. Still, it
is surprising that of the many possible ways to form a pair, so
few are actually realized. For instance, it is not imperative that
the net momentum of the pair is zero. It was predicted long ago by
Larkin and Ovchinnikov \cite{LO64} and by Fulde and Ferrel
\cite{FF64} that pairing still can occur when the electron
energies and momenta at the Fermi energy are different for the two
spin directions, for instance as the result of an exchange field
in magnetic superconductors. The resulting 'LOFF'-state is
qualitatively different from the zero-momentum state : it is
spatially inhomogeneous and the order parameter contains nodes
where the phase changes by $\pi$. The LOFF state was never
observed in bulk material, but below we present evidence that it
can be induced in a weak ferromagnet (F) sandwiched between two
superconductors (S). Such an SFS junction can yield a phase shift
of $\pi$ between the superconducting banks, as was also predicted
\cite{Bul77,Buz82,Buz92}. The $\pi$-state offers fundamentally new
ways for studying the coexistence of superconductivity and
magnetism and may also be important for superconducting
electronics, in particular in quantum computing : several schemes
for the realization of the necessary qubits (quantum two level
systems) rely on the use of phase shifts of $\pi$ in a
superconducting network \cite{Iof99,mooi99}. The spatial variation
of the superconducting order parameter in the ferromagnet arises
as a response of the Cooper pair to the energy difference between
the two spin directions in ferromagnet. The electron with the
energetically favorable spin increases its momentum by $Q$
$\propto$ $E_{ex} / v_F$, where $E_{ex}$ is the exchange energy
and $v_F$ is the Fermi velocity, while the other electron
decreases its momentum by the same amount. Since the original
momentum of each electron can be positive or negative, the total
pair momentum inside the ferromagnet is 2$Q$ or -2$Q$. Combination
of the two possibilities leads to a spatially oscillating
superconducting order parameter, $\psi(z)$, in the SFS junction
along the direction normal to the SF interfaces: $\psi(z) \propto
cos(2Qz)$ \cite{dem97,Andr91}. The same picture applies in the
diffusive limit. In this case, the oscillation is superimposed on
the decay of the order parameter due to pair breaking by
impurities in the presence of the exchange field. In the regime
$E_{ex}$ $\gg$ $k_B T$, the decay length $\xi_{F1}$ is given by
the usual expression $(\hbar D / E_{ex})^{1/2}$, where $D$ is the
electron diffusion coefficient in the ferromagnet, while the
oscillation period $2\pi \xi_{F2}$ is equal to $2\pi( \hbar
D/E_{ex})^{1/2}$. Due to the oscillations, different signs of the
order parameter can occur at the two banks of the SFS junction
when the F-layer thickness $d_F$ is of the order of half a period.
This is the so-called $\pi$-phase state, which competes for
existence with the ordinary $0$-phase state. Fig.~1a shows the
result of a Ginzburg-Landau free-energy calculation consisting of
negative condensation energy and positive gradient energy for
either state in the F-layer. One can see that the $\pi$-phase is
more favorable in the range $d_F/(2\pi \xi_{F2})$ between $0.4$
and $0.8$. Fig.~1b shows the behaviour of $\psi(z)$ in the F-layer
below and above $d_{F,cr}$, calculated using the formalism of
ref.\cite{Buz92}. The crossover from the $0$-phase to the
$\pi$-phase state should manifest itself in an anomalous thickness
dependence both of the superconducting transition temperature
$T_c$ of the junction \cite{Buz90,Rad91} and of the critical
current $I_c$ \cite{Buz82}. Experiments on the behaviour
$T_c(d_F)$ have been performed on systems such as Nb/Gd
\cite{jiang95}, Nb/Fe \cite{Muh96}, V/Fe \cite{aarts97} and Pb/Fe
\cite{Laz00} but the results are not conclusive. Especially, it
was shown that also in bilayer systems (no coupling) $T_c(d_F)$
can behave in an anomalous fashion (see \cite{Laz00} and
references therein). \\

The approach we choose is to induce the crossover as function of
temperature, not of thickness and to use a unique signature of the
junction $I_c$ : according to the Josephson relation $I_c$ =
$I_{c0} sin(\phi)$, with $\phi$ the phase difference across the
junction, biasing the junction at $\phi = \pi$ should lead to a
negative current response upon a small increase of the phase. In
other words, $I_c$ becomes negative. A change of state from $0$ to
$\pi$ will lead to a zero-crossing of $I_c$, and even if only the
absolute value of the current is measured, a sharp cusp will be
observed. The condition for having the temperature as parameter is
$k_B T$ $\approx$ $E_{ex}$. The exchange field and the temperature
then are equally important and the behaviour of the order
parameter should be written as
\begin{equation}
\psi(z )\propto e^{- z/\xi_F} \propto e^{-z/\xi_{F1}} e^{-iz/
\xi_{F2}} \;,
\end{equation}

with $\xi_F$ given by
\begin{equation}
\xi_F = \sqrt{\frac{\hbar D}{2(\pi k_B T+iE_{ex})} }  \; \; ,
\end{equation}

which yields for $\xi_{F1}$ and $\xi_{F2}$ :
\begin{equation}
\xi_{F1,2} = \sqrt{ \frac{\hbar D} {(E^2_{ex}+ (\pi k_B
T)^2)^{1/2} \pm k_B T } }  \; \; .
\end{equation}

\noindent Note that this reverts to $\xi_{F1} = \xi_{F2}$ for
$E_{ex} \gg k_B T$ as discussed above. This is the case
encountered with classical ferromagnets (Fe, Co, Ni), where
$E_{ex}$ is of the order of 1~eV and much larger than the critical
temperature $T_c$ of conventional superconductors. In the case
$k_B T \approx E_{ex}$ the decay length $\xi_{F1}$ {\it increases}
with decreasing temperature whereas $\xi_{F2}$ {\it decreases}
(see Eq. 2). This is how varying the temperature provides the
possibility to cross from a $0$-phase to a $\pi$-phase state
\cite{heikk00}. Moreover, a small value for $E_{ex}$ ensures a
large decay length $\xi_{F1}$, making it possible to fabricate
Josephson SFS sandwiches with homogeneous and continuous
ferromagnetic interlayers. Thus, the basic task is to find and
prepare such weak ferromagnets. \\

\noindent The junctions we studied consisted of superconducting Nb
(S) banks with an interlayer of a ferromagnetic Cu$_{1-x}$Ni$_x$
alloy (F). The onset of ferromagnetism in these alloys is around
$x$~= 0.44; above this concentration the Ni magnetic moment
increases with about 0.01 $\mu_B/at.\%$~Ni, which allows precise
tuning of the magnetism. An insulating SiO-layer was used between
the top electrode and the bottom SF sandwich. The window in this
layer determined the junction area of 50x50~$\mu$m$^2$. A
schematic sample cross-section is given in Fig.~2 (upper panel).
Because of the low junction resistance $R_n \approx 10^{-5}
\Omega$ the transverse transport characteristics were measured by
a SQUID picovoltmeter with a sensitivity of 10$^{-11}$~V in the
temperature range of 2.2~K to 9~K. Junctions were fabricated with
$x$ between $0.40$~and $0.57$. Upon crossing from the paramagnetic
to ferromagnetic regime the junction critical currents dropped
sharply but the $I-V$ characteristics and magnetic field
dependence $I_c(H)$ were still similar to those for standard SNS
junctions (N is a normal metal). In Fig.~2 (middle panel) $I-V$
data are shown for a junction with $x = 0.5$, $d_F$ = 14 nm at a
temperature of 4.2~K. The voltage onset at $I_c$ is sharp and well
defined. Fig.~2 (lower panel) shows that $I_c(H)$ for this
junction yields the classical 'Fraunhofer' pattern. Note that the
central peak is at zero field, even though the alloy is
ferromagnetic. This is due to the fact that the net magnetization
of the sample is zero after a careful cooldown, resulting in a
small-scale magnetic domain structure in the F-layer and zero
average phase change over the junction.   Our central result was
obtained for junctions with Cu$_{0.48}$Ni$_{0.52}$ alloys. At this
concentration the ferromagnetic transition temperature $T_{Curie}$
is about 20~K to 30~K. SQUID magnetometry for single thin films of
thickness in the range 20~nm to 100~nm showed a small hysteresis
loop close to 10~K with a coercive field of about 8~mT and a
saturation moment of 0.07~$\mu_B$/Ni~at. Fig.3 shows $I_c$(T) in
zero magnetic field for two junctions with $d_F$ = 22~nm
\cite{same}. The curve marked a) shows that $I_c$ increases with
decreasing temperature, goes through a maximum, returns to zero,
and rises again sharply. For all data points, it was ascertained
that the zero-field value was the maximum value for $I_c$. The
curve marked b) shows the same characteristic behaviour although
the zero value for $I_c$ lies at a different temperature. In this
case $I_c(H)$ characteristics were measured at three different
temperatures to ascertain that $I_c$ was determined correctly. The
data, shown in the inset of Fig.3, prove that the $I_c$(T)
oscillations are not associated with residual magnetic inductance
changes. The sharp cusp in $I_c(T)$ can be explained only by the
transition from a $0$-phase state to a $\pi$-phase state. This can
also be demonstrated by the thickness dependence of the effect.
Shown in Fig.~4a are a series of measurements for junctions of
different thicknesses in the range 23~nm to 27~nm. At 23~nm only
positive curvature is visible, an inflection point is observed for
25~nm, a maximum for 26~nm, and the full cusp now at 27~nm.
Fig.~4b shows a set of calculations based on the formalism of the
quasiclassical Usadel equations \cite{usad70}, with reasonable
parameters for $E_{ex}$ and $d_F / \xi_F^*$, where $\xi_F^* =
(\hbar D / (2\pi k_B T))^{1/2}$, demonstrating how the crossover
moves into the measurement window upon increasing the F-layer
thickness.
\\

\noindent A final remark concerns qualitative and quantitative
reproducibility. Qualitatively, the cusps can be observed for
certain thickness intervals in all sample batches with
ferromagnetic layers which are presently fabricated, both for
concentrations of 52~at.\% Ni (with $T_{Curie}$ about 20~K - 30 K)
and 57~at.\% Ni (where $T_{Curie}$ is around 100~K).
Quantitatively, there are still variations in the values of
thickness interval and temperature, as well as in the magnitude of
the critical current for different batches even with the same
nominal F-layer content. In both of these respects typical
batch-to- batch variation is demonstrated in the differences
between Figs.~2 and 3. We believe this is due to small variations
in the magnetic properties of the F-layers. In single films,
$T_{Curie}$ shows a spread of about 10~K; the weak magnetism is
apparently sensitive to the details of the preparation procedure.
\\

\noindent We thank M.Feigelman for helpful discussion and advice,
and N. S. Stepakov for assistance during the experiment. This work
was supported by a grant from the Netherlands Organization for
Scientific Research (NWO), INTAS-RFBR grant N11459 and RFBR grant
N98-02-17045.

\begin{figure}
\epsfig{file=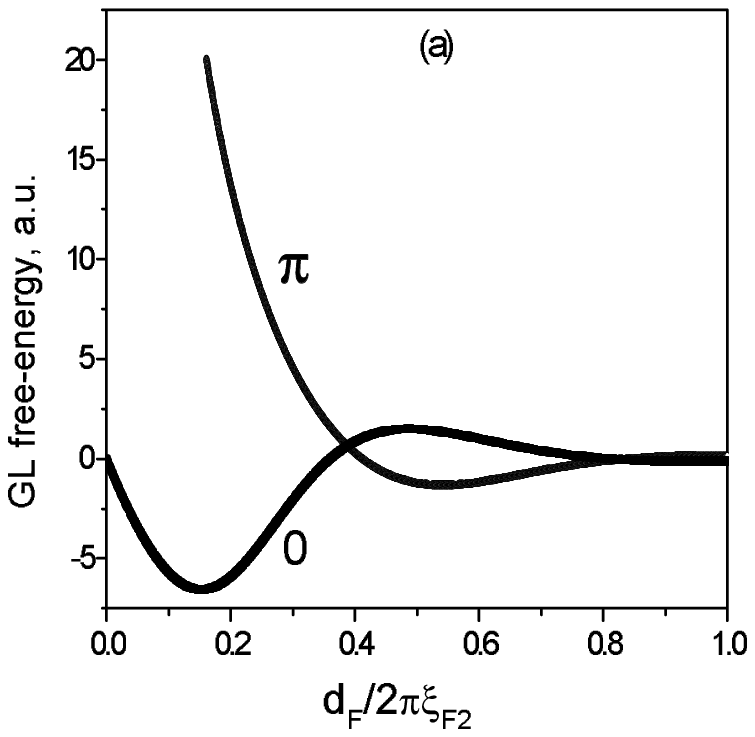,width=\columnwidth}
\epsfig{file=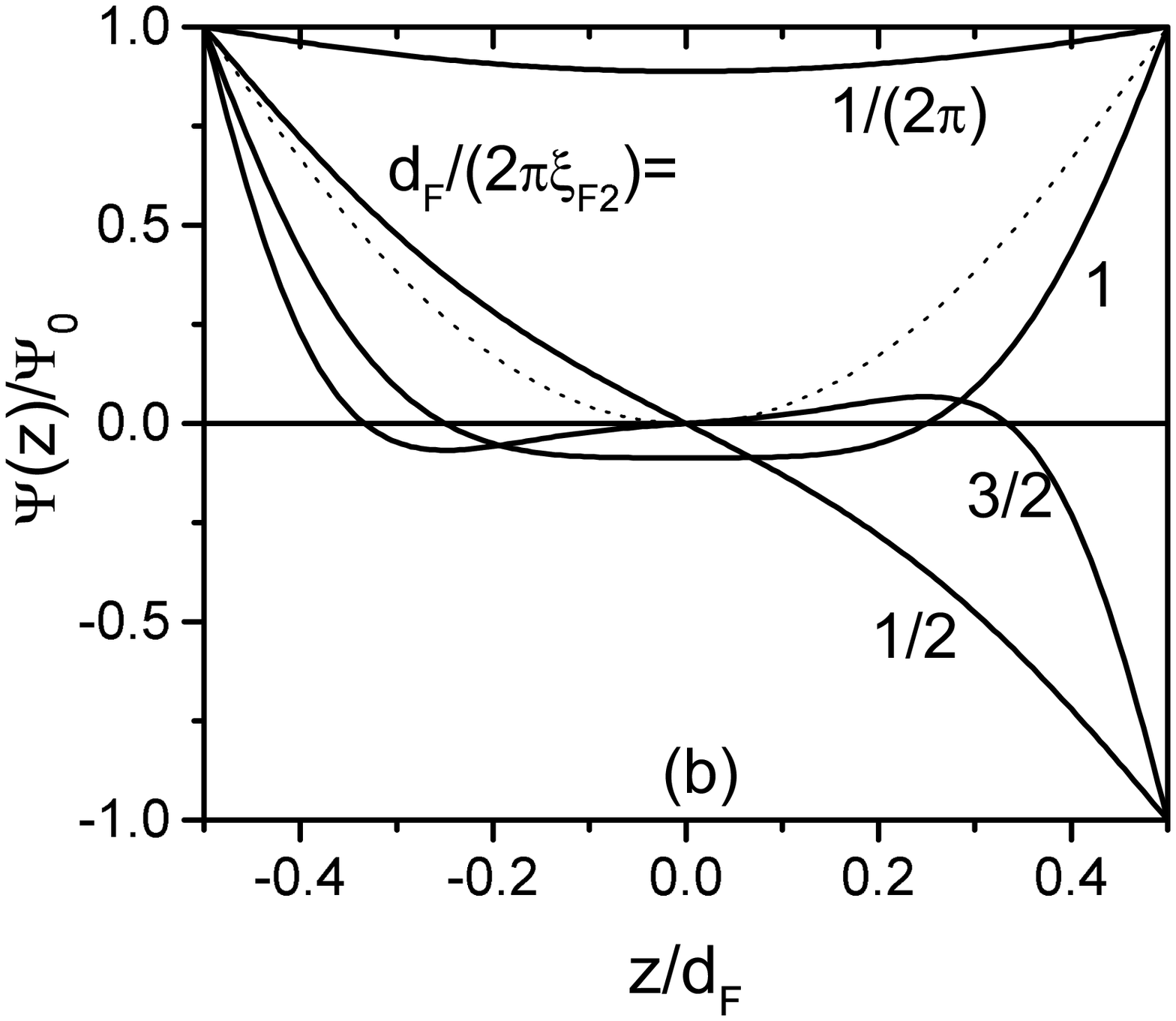,width=\columnwidth} \caption{ (a)
Calculations of the Ginzburg-Landau (GL) free energy in the
F-layer for the 0- and $\pi$-phase states. (b) The spatial
distribution of the order parameter in the F-layer of the SFS
junction calculated for various ratios of $d_F / (2\pi \xi_{F2})$
: for $d_F / (2\pi \xi_{F2})$ = 1/(2$\pi$) and 1 the lowest energy
corresponds to the 0-phase, while for $d_F / (2\pi \xi_{F2})$ =
1/2 and 3/2 the $\pi$-phase is energetically favorable. Shown for
comparison is the 0-phase for $d_F / (2\pi \xi_{F2})$ = 1/2
(dotted line), which has higher energy than the $\pi$-phase.}
\end{figure}

\begin{figure}
\epsfig{file=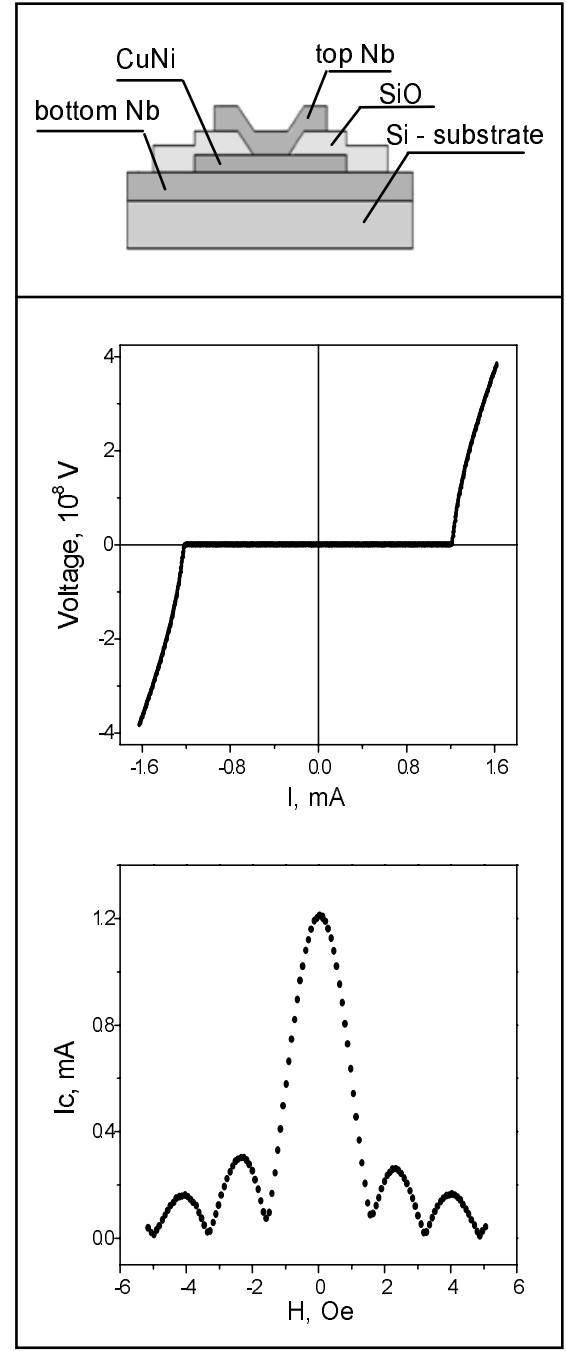,width=\columnwidth} \caption{(Upper)
Schematic cross-section of the sample. (Middle) Typical I-V
characteristic. (Lower) Magnetic field dependence of the critical
current $I_c$ for the junction with Cu$_{0.5}$Ni$_{0.5}$ and $d_F
= 14$~nm. }
\end{figure}

\begin{figure}
\epsfig{file=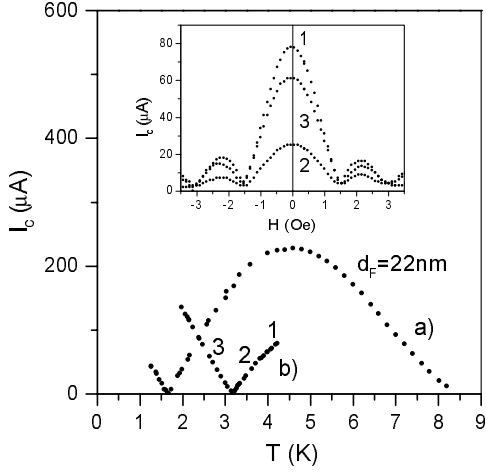,width=\columnwidth} \caption{Critical
current $I_c$ as a function of temperature $T$ for two junctions
with $Cu_{0.48}Ni_{0.52}$ and $d_F$ = 22 nm \protect\cite{same}.
The inset shows the dependence of $I_c$ on magnetic field $H$ for
the temperatures around the crossover to the $\pi$-state as
indicated on curve b : (1) T = 4.19 K, (2) T = 3.45 K, (3) T =
2.61 K.}
\end{figure}

\begin{figure}
\epsfig{file=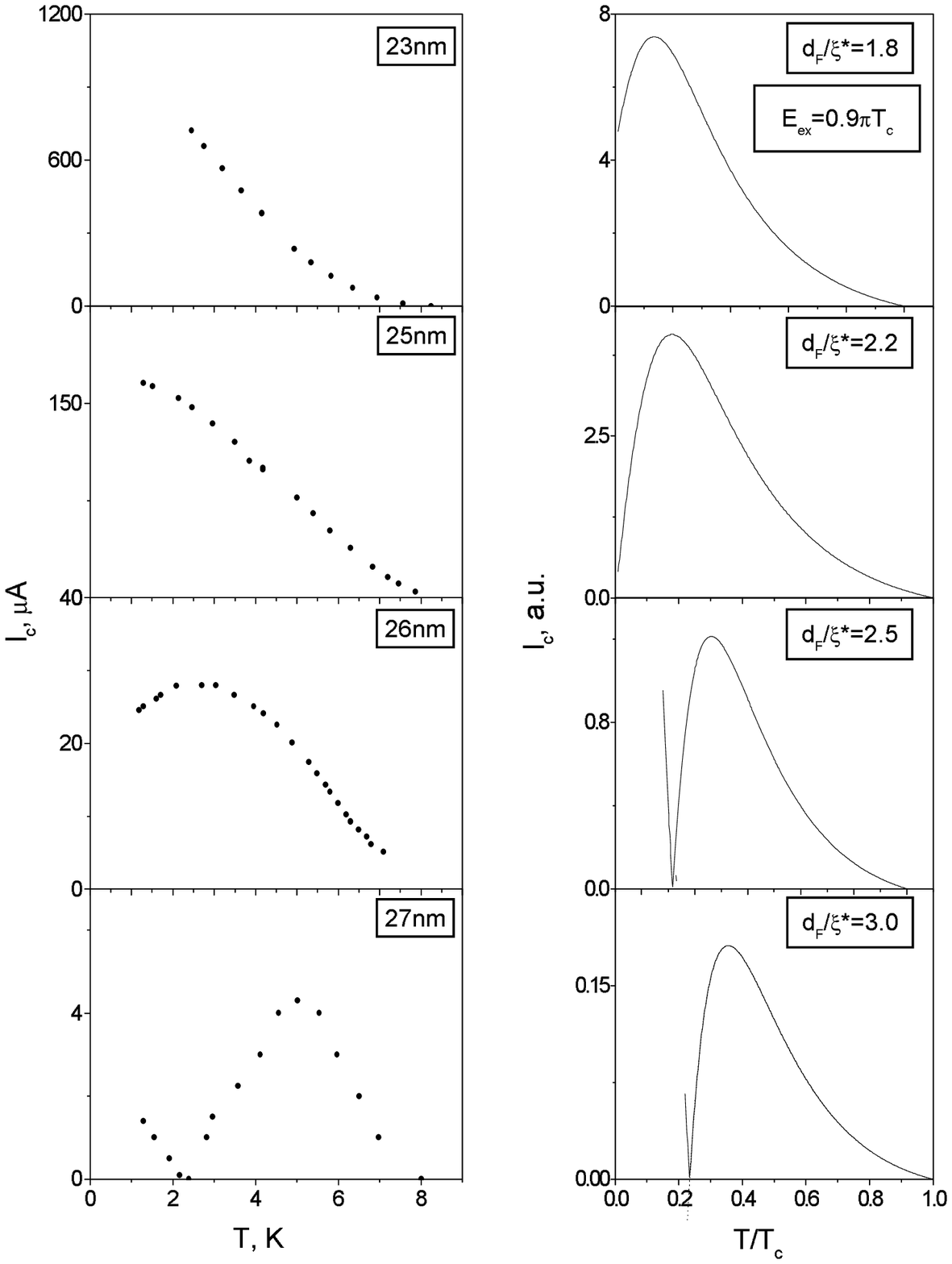,width=\columnwidth} \caption{Left : critical
current $I_c$ as function of temperature for $Cu_{0.48}Ni_{0.52}$
junctions with different F-layer thicknesses between 23~nm and
27~nm as indicated. Right : model calculations of the temperature
dependence of the critical current in an SFS junction for $E_{ex}
= 0.9 \pi T_c$ and various ratios of $d_F/\xi^*$, where $\xi^*$ =
$\sqrt{\hbar D / (2 \pi k_B T_c) }$ }
\end{figure}


\begin{thebibliography}{99}
\bibitem{same} It was actually the same junction, but removed after
the first measurement between 4.5 K and 2 K and contacted again.
We assume that the heating during the contacting procedure led to
small changes of the F-layer properties.
\end{thebibliography}
\end{document}